# VANISHING OF ELECTRON-HOLE ASYMMETRY IN NANO-SIZED CHARGE ORDERED MANGANITES $Pr_{(1-x)}Ca_{(x)}MnO_3$ (x=0.36, 0.64) – EPR STUDIES

*K.G.Padmalekha and S.V.Bhat*

*Department of Physics, Indian Institute of Science, Bangalore-560012.*

**Abstract**

We observe a disappearance of electron-hole asymmetry in hole and electron doped manganites of composition $Pr_{(1-x)}Ca_{(x)}MnO_3$ (x=0.36, 0.64), as reflected in the g-parameter values. The bulk property of opposite g-shifts in room temperature seems to have disappeared in nanoparticles. We analyze the results in the light of ferromagnetic fluctuations and melting of charge order in nanoparticles.

## I. INTRODUCTION

Properties of the hole-doped manganites ($x < 0.5$) when compared with those of the electron-doped ($x > 0.5$) compositions show some marked differences. Holes are the majority carriers in the former case, whereas, electrons are the majority careers in the latter. Charge ordering is the dominant interaction in the latter class of manganites unlike ferromagnetism and metallicity in the hole-doped materials.[1] Electron-hole asymmetry in the PCMO system has been studied in detail in the single crystalline and polycrystalline form using the EPR parameters[2], specially the 'g' values, as well as thin film form.[3] The two systems studied were $Pr_{0.64}Ca_{0.36}MnO_3$ (PCMH) – hole doped and $Pr_{0.36}Ca_{0.64}MnO_3$ (PCME) – electron doped.

Both bulk PCMH and bulk PCME get charge ordered in the paramagnetic state, with $T_{CO}$ of 210 K and 268 K, respectively.[4] They show maxima in the magnetization curves at the charge-ordering transition temperatures. Bulk PCMH shows an AFM transition around 140 K whereas bulk PCME does not show an AFM transition. Both PCMH and PCME are insulators down to low temperatures, as is expected of charge-ordered compositions, but PCME shows a more marked change in resistivity at $T_{CO}$ than PCMH. The difference between the two lies in the effect of magnetic fields.[4] Charge ordering in PCMH can be collapsed by a high enough magnetic field like 12 T [5] and destabilized by electric field[6] even inducing a metal-insulator transition.[7] Application of pressure also has similar effect[8] whereas, a 12 T magnetic field has no effect whatsoever on the resistivity of PCME.[4] On doping with 3% $Cr^{3+}$, PCMH becomes ferromagnetic with a $T_C$ of 130 K, but PCME remains paramagnetic and charge-ordered, albeit with a slightly lower $T_{CO}$ (215 K). Doping with 3% Ru shows similar differences between the two manganites.

PCMH in bulk form shows ferromagnetic fluctuations above $T_{CO}$.[9, 10] These fluctuations weaken and give way to antiferromagnetic fluctuations below $T_{CO}$ and ferromagnetic fluctuations completely disappear below the $T_N$ = 170 K. Below $T_N$, charge exchange (CE) type of antiferromagnetic order is established [11, 12]. A long range orbital order is never established in PCMH and the correlation length of orbital ordering is found to be much smaller than the correlation length of charge ordering [13].

It is possible that even in case of PCME bulk, ferromagnetic fluctuations are present just like PCMH in the paramagnetic phase [4, 12]. These fluctuations decrease below $T_{CO}$ giving rise to antiferromagnetism, probably of CE type. Theoretically, complete orbital order is considered to be less likely in electron doped manganites [14].

The thin film study by Parashar et.al.,[3] showed that the charge ordering in PCMH could be melted by the application of magnetic field of 5 T whereas the charge ordering in PCME does not melt. The effect of electric field however, is similar in both the cases though the metal-insulator transition is more pronounced in the case of PCMH. For both $Mn^{3+}$ and $Mn^{4+}$ ‘g’ (the Lande g-factor) is expected to be lower than the free electron g-value, denoted by $g_e$ which is equal to 2.0023. In the octahedral crystal field of Oxygen atoms, the g-value for $Mn^{3+}$ is supposed to be 1.97 and for $Mn^{4+}$ it is 1.99 [15]. However, the study by Joshi et.al., [2] showed that in polycrystalline samples, ‘g’ parameter was found to be shifted in opposite directions at room temperature.

The shift in g-value is given by the equation [15]:

$$g = g_e\left(1 - k\frac{\lambda}{\Delta}\right) \qquad (1)$$

where, k is a positive numerical factor, $\lambda$ is the spin-orbit coupling constant and $\Delta$ is the crystal field splitting.

The anomalous ‘g’ shift was attributed to the ‘hole’ nature of the charge carriers in PCMH since for holes in a less than half filled shell the spin–orbit coupling constant $\lambda$ is negative [16]. The increase in ‘g’ below $T_{CO}$ found in PCMH was attributed to the strengthening of the spin–orbit interaction and spin–other orbit interaction due to orbital ordering developing between $T_{CO}$ and $T_N$ [2, 17].

In this paper, we report EPR studies on PCMH and PCME nanoparticles showing vanishing electron-hole asymmetry. This is reflected in EPR g-parameter and its temperature dependence.

## II. EXPERIMENTAL METHODS

Nanoparticles of PCMH and PCME were prepared using the sol-gel method[18]. Stoichiometric amounts of nitrates of praseodymium, calcium and manganese were mixed in a solvent containing equal amounts of ethylene glycol and water. The mixture was heated with stirring until a thick sol was formed. After the solvent had evaporated, the resulting resin was finely ground and annealed at 650 $^0$C to obtain the nanoparticles of PCMH and PCME. XRD, TEM and SQUID data analysis were done on these nanoparticles. The XRD results along with Rietveld analysis are given in Fig.1a for PCMH and Fig. 1b for PCME nanoparticles. TEM images for nano PCMH (Fig 2a) and nano PCME (Fig. 2c) with the insets (Fig. 2b and Fig. 2d respectively) show the size distribution.

SQUID magnetization data for nano PCMH is shown in Fig. 3. The magnetization data for nano PCME is shown in Fig. 4.

EPR measurements were carried out using a Bruker EMX X-band EPR spectrometer on nano PCMH and nano PCME dispersed in poly vinyl alcohol.

Temperature was varied from 3.8 K to 300 K. The signals were Lorentzian in shape and signals were observed down to 30 K in the case of PCMH and down to 60 K in the case of PCME. A speck of DPPH was used as a field marker to enable accurate determination of the g-parameter. The signals were fit to the Lorentzian equation given by equation 2.

$$\frac{dP}{dH} = \frac{d}{dH} A.\left[ \frac{\Delta H}{4(H - H_0)^2 + \Delta H^2} + \frac{\Delta H}{4(H + H_0)^2 + \Delta H^2} \right] \tag{2}$$

where, $\Delta H$ is the full width at half intensity, $H_0$ is the resonance field and A is a scaling parameter proportional to the intensity. Some fitted signals for both PCMH and PCME are shown in Fig. 6. The fitting parameters obtained from the fitting are plotted in Fig. 7. The linewidth that has been plotted is peak to peak linewidth which is related to $\Delta H$ by the relationship $\Delta H_{PP} = \Delta H/\sqrt{3}$. The intensity multiplied by temperature of the signals is plotted in Fig. 8 for both PCMH and PCME. The resonance field shows a monotonous decrease for both the samples as the temperature decreased.

The 'g' value for DPPH was taken as 2.0036. The calculated g-values are plotted in Fig. 7. The high temperature parts of the 'g' values are given in the inset of Fig. 7 comparing them with the free electron g-value ($g_e = 2.0023$). The relationship between resonance field and 'g' value is given by the equation:

$$h\nu = g\beta H_0 \tag{3}$$

where, h is planck's constant, $\nu$ is the measuring frequency, $H_0$ is the resonant magnetic field, $\beta$ is the Bohr magneton and g is the Lande's g factor.

## III. RESULTS AND DISCUSSION

### A. XRD and TEM results

The TEM results show that the nanoparticles are crystalline with a fair degree of size distribution. The XRD patterns showed single phase had formed in both the cases of PCMH and PCME. Rietveld analysis of the data showed that the lattice parameters had increased from bulk to nano with an increase in the Mn-O-Mn bond angle (table 1). The size of the nano crystallites was calculated using the Scherrer formula. The size of PCMH was found to be ~ 20 nm and for PCME ~ 16 nm. The TEM measurement showed higher average particle size since the XRD measurements give the crystallite size and TEM measurements give the particle size.

### B. Magnetization

Magnetization data from Fig. 3a shows that the charge order has completely disappeared in nano PCMH showing no $T_{CO}$ in the entire temperature range with magnetization monotonically increasing with decreasing temperature and we see that at $T < 100$ K, ferromagnetism picks up and hysteresis loop opens up (Fig. 3b). It has been observed that in charge ordered manganites, size reduction melts the charge order and induces ferromagnetism.[18] $T_C$ for nano PCMH is ~ 55 K. Magnetization data for nano PCME (Fig .4a) shows that the charge order peak has decreased in intensity but it still is discernible at ~ 220 K. We see that here also, ferromagnetism picks up at $T <100$ K and hysteresis loop opens up (Fig. 4b). It has been established by the bulk studies that PCME is a more robust charge ordered system and melting the charge order is difficult in it. The size reduction shows similar tendency. The magnetization per gram for nano PCMH being an order of magnitude higher than that of nano PCME at a similar temperature suggests that FM is favoured more in the case of nano PCMH as it is closer to the metallic regime where ferromagnetic double exchange is possible.

At the lowest temperature of measurement, (1.9 K), we also see an exchange bias in both these systems (Fig. 5a and 5b). Exchange bias is usually defined as $H_{EB} = |H_1 + H_2| / 2$ and the coercive field is defined as $H_C = |H_1 - H_2| / 2$ where, $H_1$ and $H_2$ are the left and right coercive fields respectively.[19] For nano PCMH, at 1.9 K, $H_{EB}$ is found to be 111.95 Oe and $H_C$ is found to be 2079.08 Oe. For nano PCME, $H_{EB}$ = 733.01 Oe and $H_C$ = 2372.28 Oe. The presence of this exchange bias suggests the co-existence of FM and AFM regions in the sample. Exchange bias in charge ordered nanoparticles has been explained by a core-shell model[20] in which the AFM core is surrounded by a FM shell and the pinning of the FM spins by the AFM spins is responsible for the exchange bias.

**C. EPR parameters**

Now we attempt at giving an explanation for the observed behavior of different EPR parameters.

We see that unlike for the bulk case, the 'g' parameters for both the nano samples lay on the same side of the free electron 'g' value. We also see that their behavior as the temperature reduces is also similar as shown in Fig.7. This we term as the "disappearance of electron-hole asymmetry" as seen in EPR 'g' parameter values. We see that in case of PCME nanoparticles, the 'g' value decreases as the particle size decreases. This is given in Table 2. We see that the difference reduces monotonically as the size of the particle reduces.

The PCMH 'g' value shifts to a higher value than the bulk 'g' value. This can be explained by considering the ferromagnetic fluctuations which are known to be present in the system of PCMH as said earlier. Jirak et.al.,[12] point out that ferromagnetic fluctuations exist in the PCMH composition and the coupling along the (001) direction in the crystal is ferromagnetic. They also propose a two-phase model consisting of FM and AFM domains in the system. It has been shown that as the size reduces, charge order completely disappears in charge ordered manganites in half-doped PCMO, [21, 22] similar to our observation in PCMH nanoparticles. Non- zero magnetic moments have been observed in nanoparticles of systems which were antiferromagnetic in their bulk form [23]. We can say ferromagnetism is favoured in the PCMH system and we observe the magnetization behavior accordingly. The ferromagnetism in the nanoparticles can push the 'g' value to higher values, with apparent shift in resonance field because of the intrinsic magnetization of the ferromagnetic domains.

In the case of nano PCME, ferromagnetic fluctuations are known to be present[12], which can explain the magnetization behavior. Though we do not see complete disappearance of CO as in the nano PCMH case, the strong reduction in the charge ordered peak in the magnetization data means that ferromagnetic fluctuations are getting stronger. Similar effect of internal field as in the case of nano PCMH due to ferromagnetism can shift the g-value to higher values.

We see from our Rietveld analysis (table 1) that the lattice parameters are larger in the cases of both nano PCMH and nano PCME than the bulk values. We also see that that Mn-O-Mn bond angle has increased enabling better ferromagnetic interactions in the system.

The EPR linewidths (Fig.8) are known to have contribution from the anisotropic crystal field effects and the Dzyloshinsky-Moriya exchange interactions. Though the reported linewidths in case of bulk PCMH was less than that of bulk PCME at room temperature, we see that in the case of nanoparticles, the linewidth has opposite behavior at room temperature. The linewidth in case of PCMH behaves similar to that of CMR manganites, i.e., showing a change in slope near ~ 1.1 $T_C$.[24, 25] The $T_C$ in this system is very low and far from $T_C$, the effect on linewidth can get randomized resulting in the different linewidth values at room temperature. In PCME, we see a similar increase in linewidth as temperature reduces.

The intensity multiplied by temperature plot shows very different behavior in nano PCMH (Fig. 9a) and PCME (Fig. 9b). PCMH shows increased intensity as temperature decreases, as expected in metallic and ferromagnetic systems. The reduction in intensity at lower temperature is because part of the signal is getting lost because resonance field has moved significantly to lower values. In case of PCME, there are competing effects of CO, AFM and FM in the system. There can be two possible scenarios in the system: i) The system can be intrinsically charge ordered with both AFM and FM coexisting in each nanoparticle. The maxima near $T_{CO}$ is commonly seen in charge ordered manganites. These two competing intereaction within the same nanoparticles can explain the variation in intensity. ii) The size distribution form TEM images show that there is a wide variety of sizes in the system. The larger ones still showing AFM at low temperatures are responsible for the reduction in intensity below $T_{CO}$ until FM signal picks up from the smaller sized ones which are fully ferromagnetic giving us the increasing intensity at lower temperatures.

As can be seen from table 2, increasing particle size tends to take the g-value more towards the free-electron g-value, finally crossing it and going to the bulk PCME g-value. This effect also corroborates the hypothesis that decreasing ferromagnetism as size increases reduces the g-value.

If we consider size-induced metallicity in the system, then like in the case of other more-metallic CMR manganites like LCMO [26], metallicity tends to average out the anisotropies in the system resulting in reduction in the asymmetry across half-doping.

In nano PCMH, charge ordering has completely disappeared whereas we see a large reduction in charge ordering peak in magnetization value for nano PCME. This reduction is proportional to the size of the particle. The slight difference in different parameters that we are observing in the nano PCMH and PCME could be attributed to the residual charge ordering still present in PCME. If one is able to completely melt the charge ordering in PCME by reducing the particle size sufficiently, one maybe able to notice a complete disappearance of electron-hole asymmetry in the PCMO system.

## IV. CONCLUSIONS

We report a disappearance of electron-hole asymmetry in the nanoparticles of PCMO on either side of the half-doping regime as seen in the EPR 'g'-parameter. We analyze the results in the light of increasing ferromagnetic fluctuations as the particle size reduces. The changes in Mn-O-Mn bond angles and possible increase in metallic behavior is considered as another possible explanation for this behavior. Increasing the size of the nanoparticles is seen to push the PCME system towards the free electron g-value, suggesting that the size reduction is responsible for the disappearance of electron-hole asymmetry.

## ACKNOWLEDGEMENTS

The authors would like to acknowledge Professor Arun Grover for the magnetization experiments. KGPL would like thank M.M.Borgohain for help in doing EPR measurements. The funding for the work was provided under the Nanoscience and Technology Initiative of the Nano Mission, DST to SVB. Useful discussion with V. B. Shenoy is gratefully acknowledged

**FIGURE CAPTIONS**

**Figure 1:** XRD and Rietveld analysis of (a) nano PCMH (b) nano PCME.
The solid dots are experimental signal and the solid line is the fit.
The dotted line is the difference.

**Figure 2:** (a) TEM image of nano PCMH particles with (b) size distribution. (c) TEM image of nano PCME particles with (d) size distribution.

**Figure 3:** (a) Magnetization data for nano PCMH. (b) ferromagnetic hysteresis loop at 1.9 K.

**Figure 4:** (a) Magnetization data for nano PCME. (b) ferromagnetic hysteresis loop at 1.9 K.

**Figure 5:** (a) Exchange bias for nano PCMH and (b) nano PCME at 1.9 K.

**Figure 6:** Fitted Lorentzian lineshapes for (a) nano PCMH and (b) nano PCME, at different temperatures. The Solid lines are the experimental curve and the dots represent the fit. The sharp line in the centre is due to DPPH.

**Figure 7:** Calculated g values from the fit to equation 2. The hollow circles are for nano PCMH and solid circles are for nano PCME.

**Figure 8:** EPR linewidth as a function of temperature. The hollow circles are for nano PCMH and solid circles are for nano PCME.

**Figure 9:** EPR intensity * temperature as a function of temperature for (a) nano PCMH and (b) nano PCME.

**Table 1:** Lattice parameters and bond angles for PCMH and PCME bulk and nano.

**Table 2:** Variation of 'g' value at 300 K as the size of PCME nanocrystals increase.

**Figure 1**

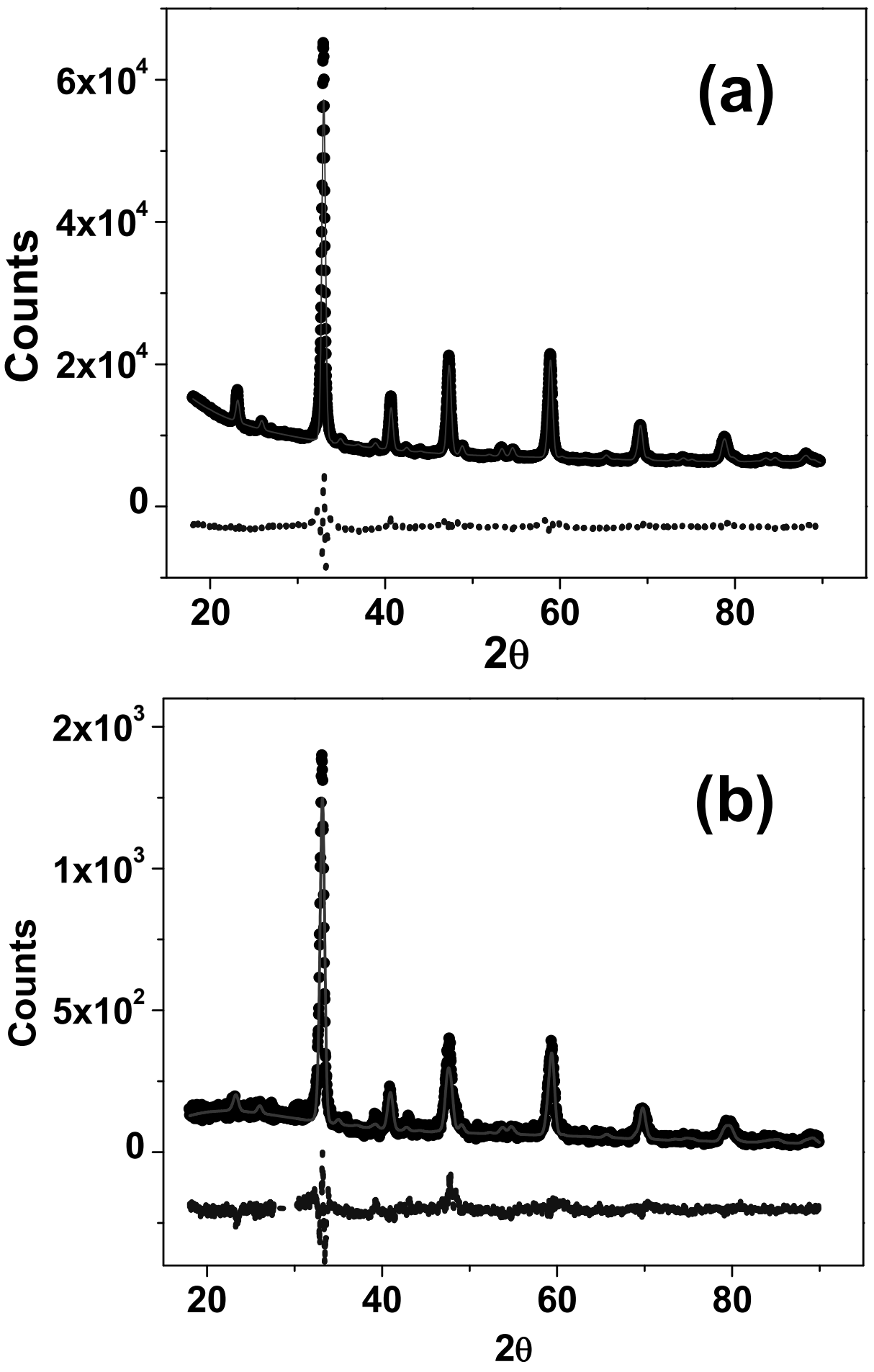

K.G.Padmalekha and S.V. Bhat

**Figure 2**

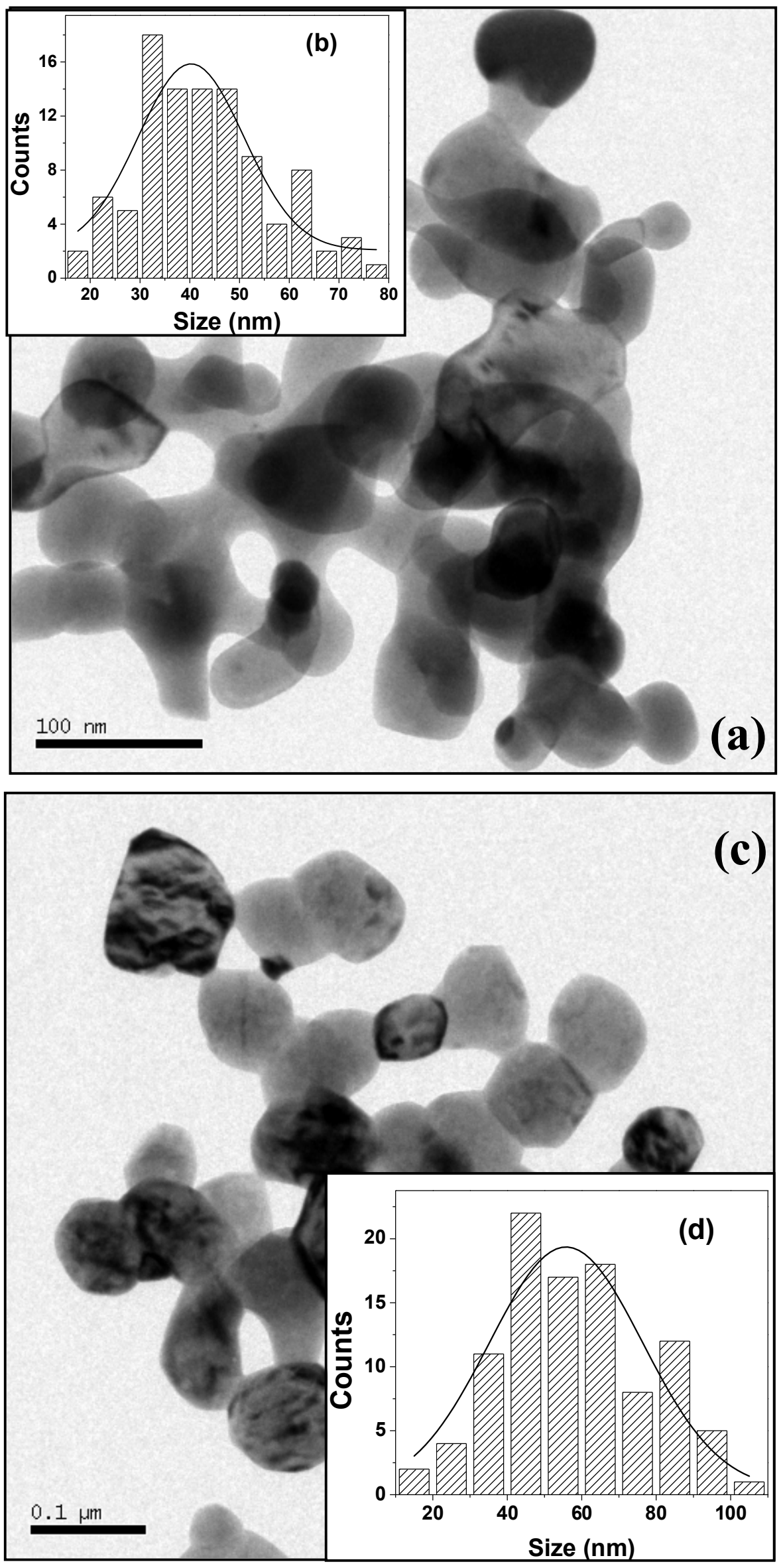

K.G.Padmalekha and S.V. Bhat

**Figure 3**

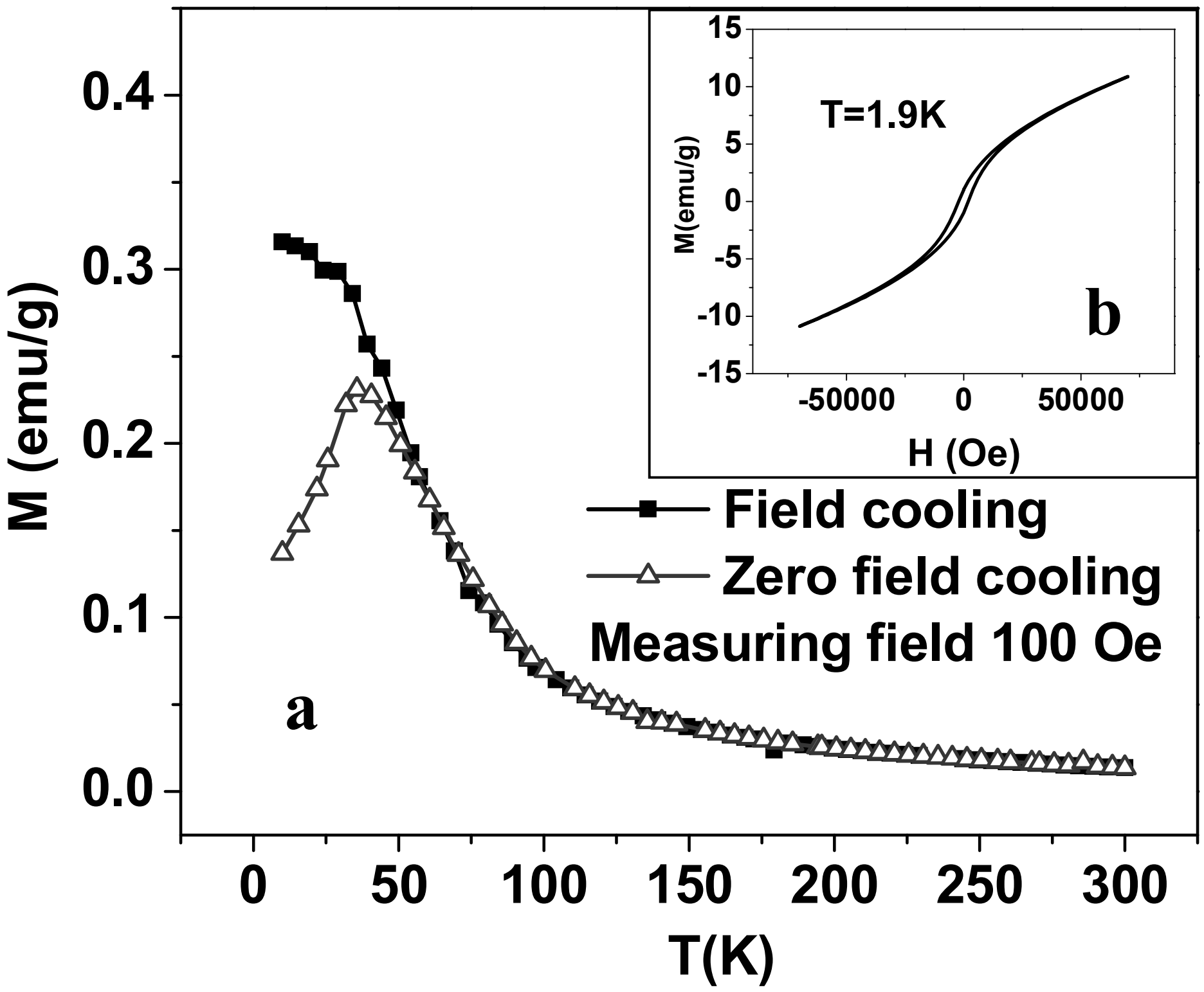

K.G.Padmalekha and S.V. Bhat

**Figure 4**

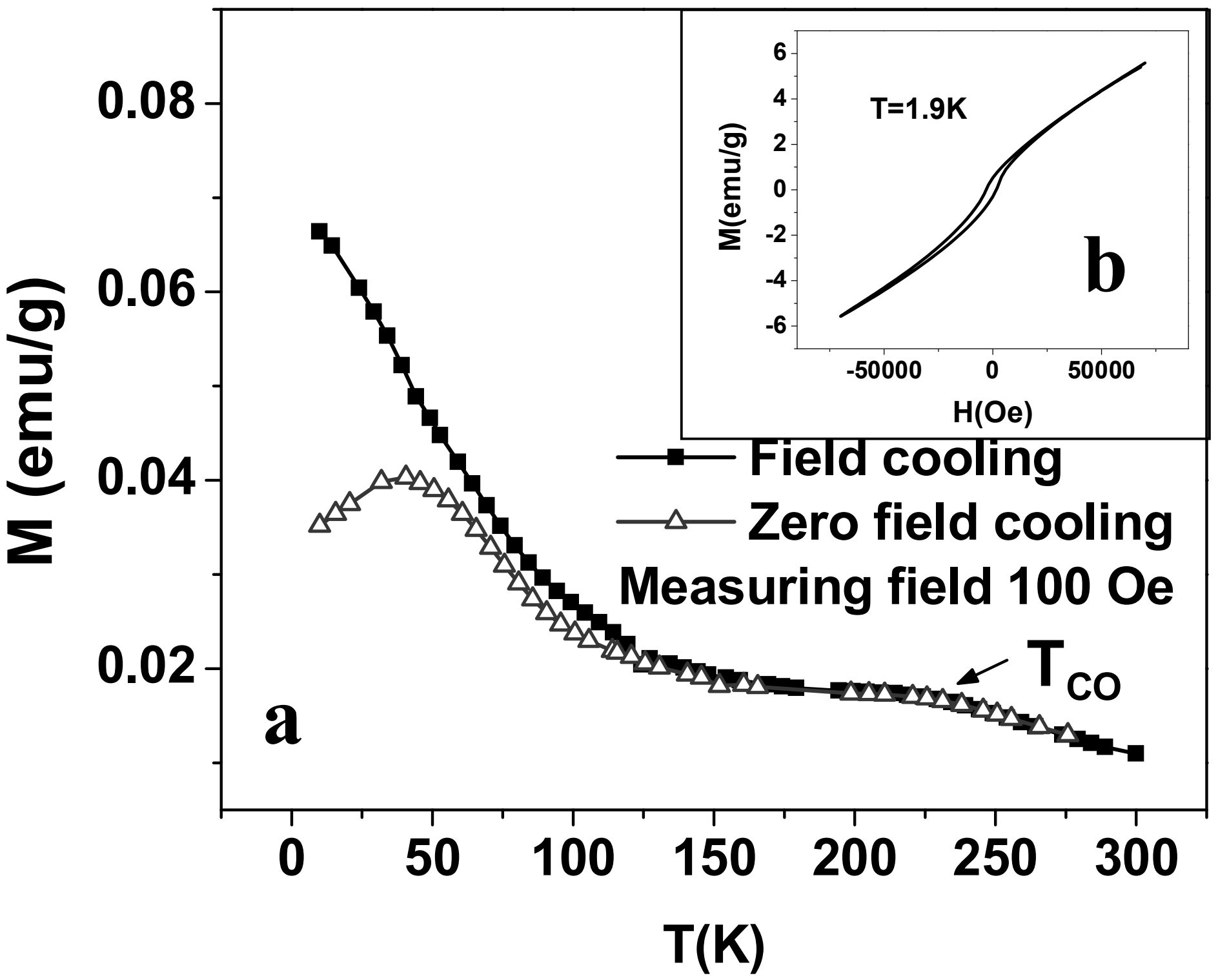

K.G.Padmalekha and S.V. Bhat

**Figure 5**

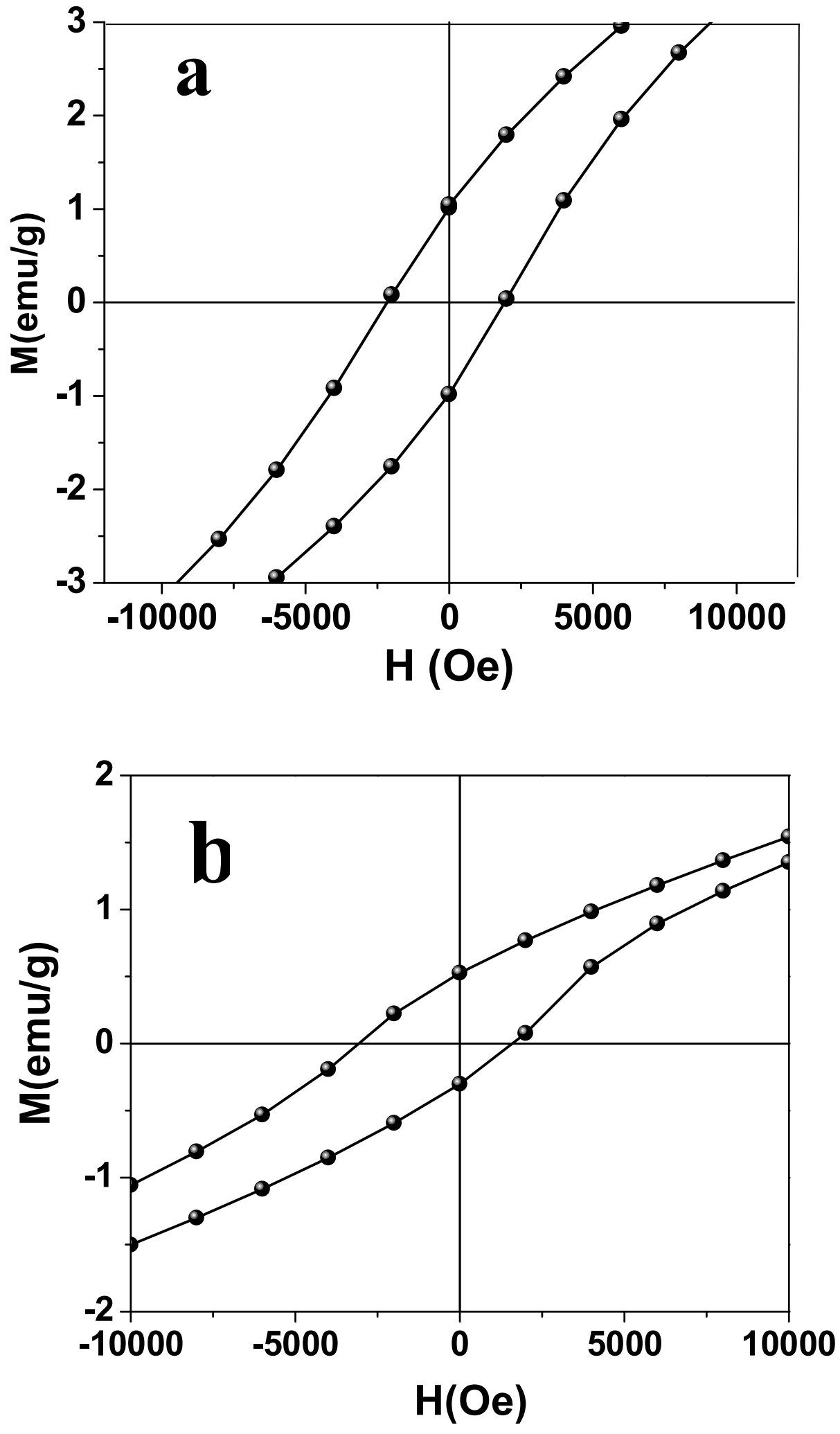

K.G.Padmalekha and S.V. Bhat

**Figure 6**

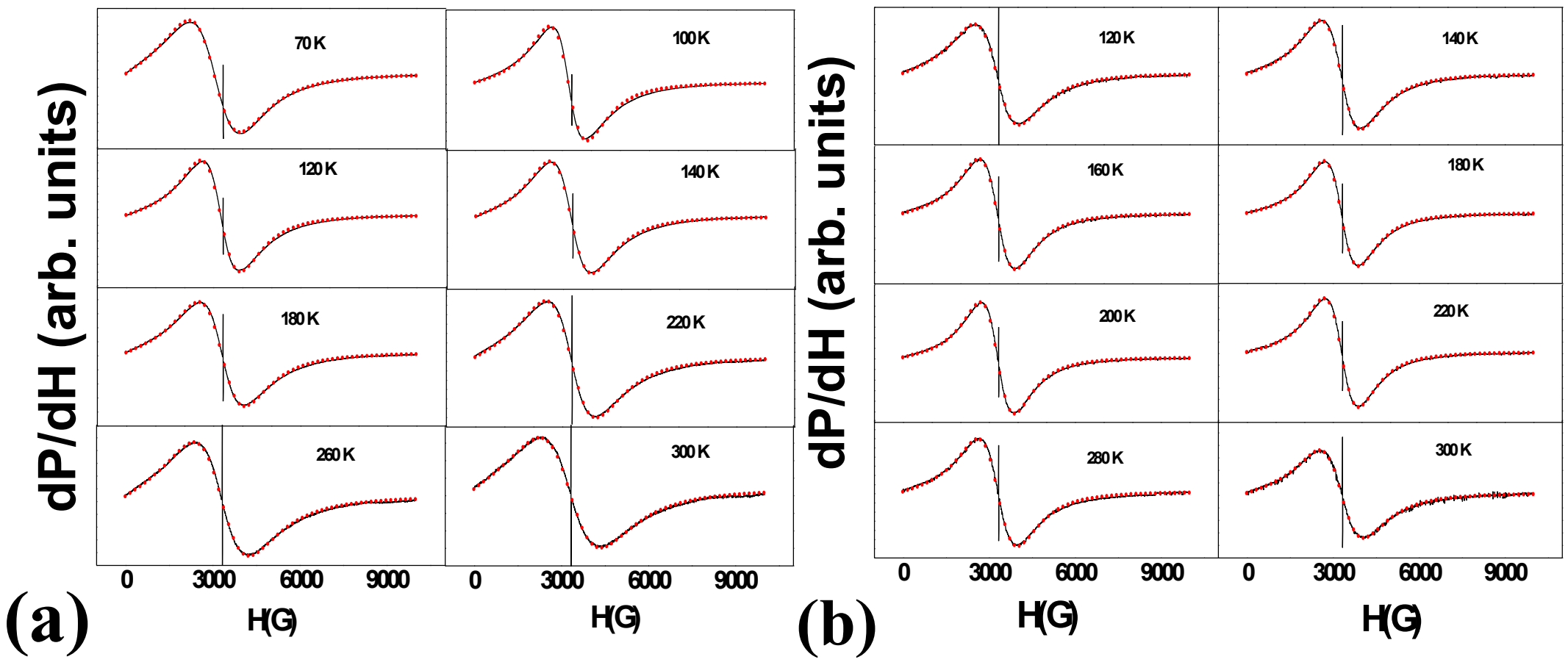

K.G.Padmalekha and S.V. Bhat

**Figure 7**

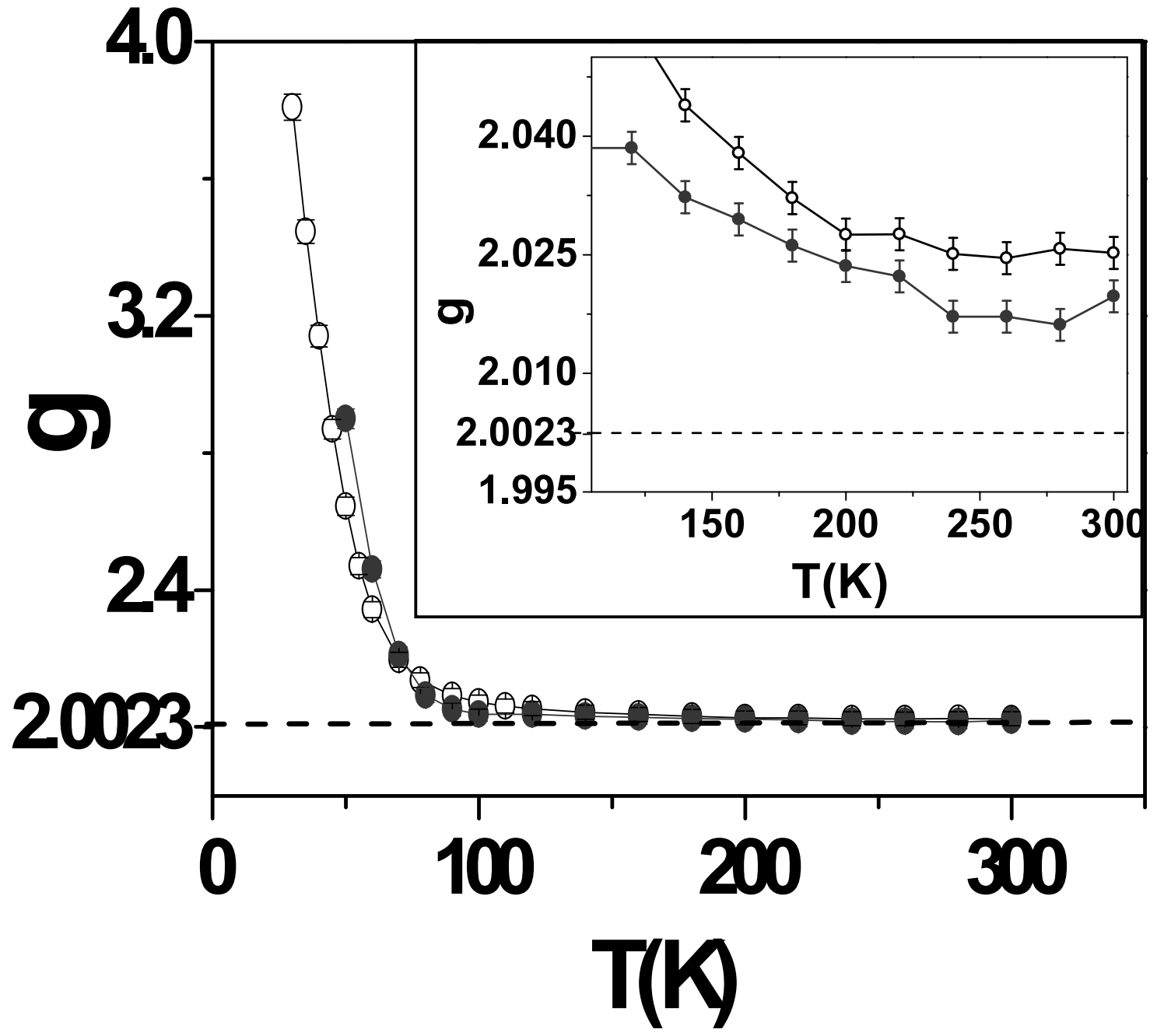

K.G.Padmalekha and S.V. Bhat

**Figure 8**

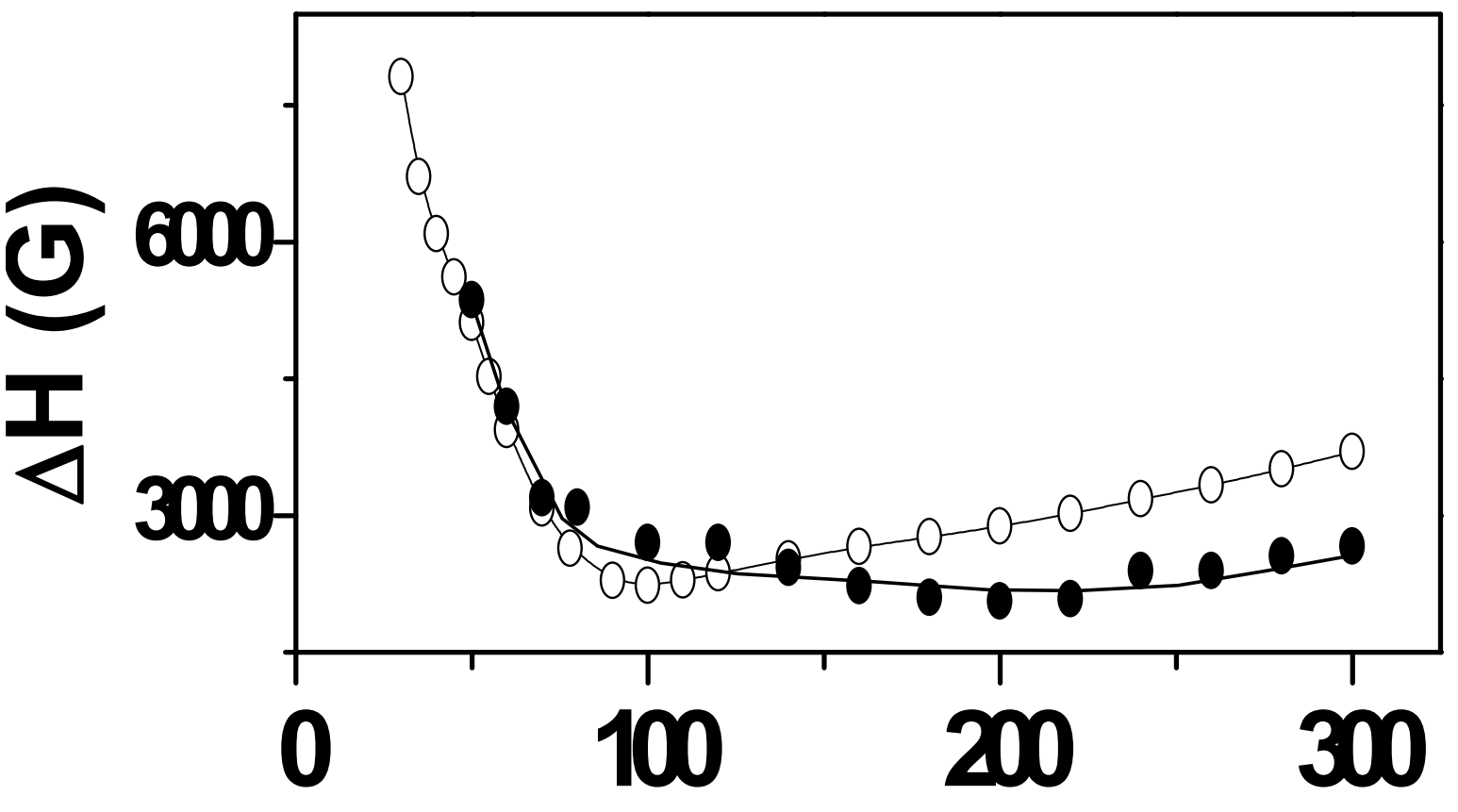

K.G.Padmalekha and S.V. Bhat

**Figure 9**

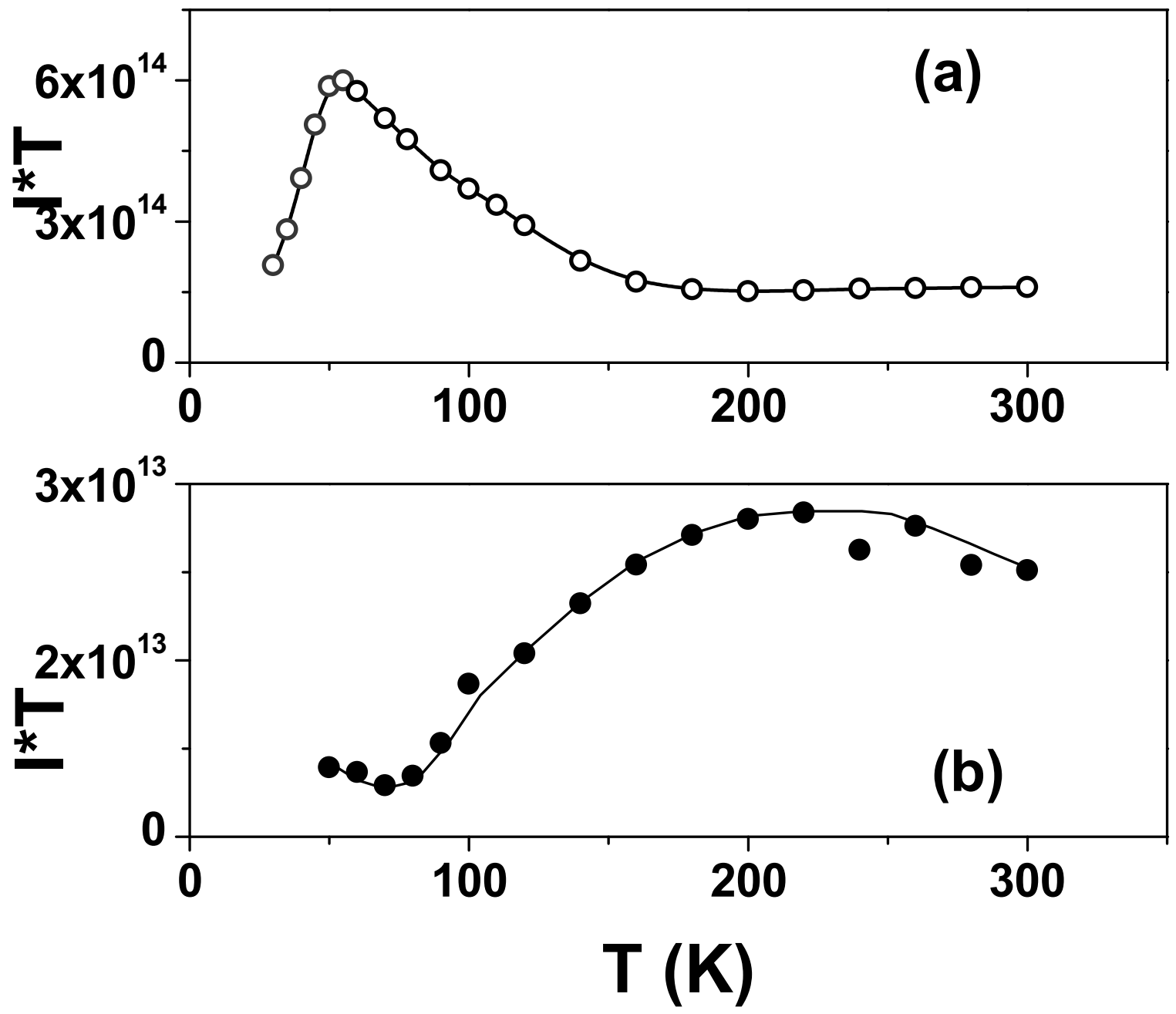

K.G.Padmalekha and S.V. Bhat

**Table 1**

| Sample | Lattice parameters (Space group = Pbnm, $\alpha$=$\beta$=$\gamma$=$90^0$) | | | Mn-O-Mn bond angle | $W_{RP}$ |
|---|---|---|---|---|---|
| | a | b | c | | |
| PCMH bulk | 5.41976 | 5.43966 | 7.65646 | $157.486^0$ | 4.8 % |
| Nano PCMH (~ 20 nm) | 5.42838 | 5.45588 | 7.66218 | $157.502^0$ | 4.32 % |
| PCME bulk | 5.37575 | 5.36760 | 7.56604 | $157.499^0$ | 5.7 % |
| Nano PCME (~ 20 nm) | 5.37849 | 5.41827 | 7.58019 | $157.535^0$ | 4.89 % |
| Nano PCME (~ 16 nm) | 5.38562 | 5.43563 | 7.58678 | $157.551^0$ | 9.29 % |

**Table 2**

| Size of PCME (calculated from XRD) | 'g'- value at 300 K |
|---|---|
| 16 nm | 2.0197 |
| 20 nm | 2.0180 |
| 45 nm | 2.0071 |
| Bulk | 1.9832 |